\documentclass{article}

\usepackage{microtype}
\usepackage{graphicx}
\usepackage{hyperref}

\usepackage[accepted]{mlsys2025}

\makeatletter
\renewcommand{\Notice@String}{}
\makeatother

\usepackage{balance}
\usepackage{subcaption}
\usepackage{xspace}
\usepackage{enumitem}
\usepackage{booktabs}
\usepackage{tabularx}
\usepackage{array}
\usepackage[most]{tcolorbox}

\newcommand{\finding}[2]{\par\vspace{1ex}\noindent\textbf{Finding #1: }\textit{#2}\par}
\newcommand{\subfinding}[2]{\par\vspace{1ex}\noindent\textbf{Subfinding #1: }\textit{#2}\enspace}
\newcommand{\storageimplication}[1]{\par\smallskip\noindent\textbf{Storage implication.}\enspace #1\par}

\newcommand{\stitle}[1]{\vspace{1ex}\noindent\textbf{#1}}

\mlsystitlerunning{The Storage View of KV Cache}

\begin{document}

\twocolumn[
\mlsystitle{The Storage View of KV Cache:\\
Characterizing Cache Behavior Across LLM Workloads}

\mlsyssetsymbol{intern}{$\dagger$}
\begin{mlsysauthorlist}
\mlsysauthor{Talia Q. Chen}{utaustin,intern}
\mlsysauthor{Daniel Waddington}{ibm}
\mlsysauthor{Narangerelt Batsoyol}{ibm}
\mlsysauthor{Vijay Chidambaram}{utaustin}
\end{mlsysauthorlist}
\mlsysaffiliation{utaustin}{The University of Texas at Austin}
\mlsysaffiliation{ibm}{IBM Research}

\mlsyscorrespondingauthor{Talia Q. Chen}{taliac@utexas.edu}

\mlsyskeywords{KV cache, LLM serving, storage systems, workload characterization}

\vskip 0.3in

\begin{abstract}
Modern LLM serving systems increasingly extend KV caches beyond GPU memory
into host DRAM, SSDs, and remote storage, yet these stores are largely managed
using workload-agnostic policies such as LRU and fixed TTLs.
To guide the design of KV storage for LLMs, we ask what \emph{execution
context} storage needs to make retention, reclamation, and placement decisions
across heterogeneous LLM workloads.

We characterize KV-cache reuse, lifetime, and footprint at the granularity of
KV-cache blocks across \textbf{11 production and public traces spanning
six categories}: chat, API serving, RAG, coding agents, tool-using agents, and
multi-lineage workflows.
We analyze where reuse comes from, when and how frequently KV blocks recur, how
long they remain useful, and how they use storage capacity.
Across workloads, KV state exhibits two recurring forms.
\emph{Single-session state} is reused within one execution: its footprint grows
as sessions deepen and accumulates across concurrent sessions, and it can remain
useful through long idle periods but loses value when the session completes.
\emph{Multi-session state} recurs across sessions of the same workload, outlives
individual sessions, and ranges from hot prefixes to a large infrequent tail.

While prior systems often exploit application-, agent-, or workflow-specific
information, our results suggest that much of the structure needed for
KV-storage management can be exposed through
three generic signals: a workload identifier, a session
identifier, and a session completion signal.
Trace-driven comparisons quantify their value: workload-specific recurrence
estimates 
retain up to 22\% more within-session hits and reduce dead state by up to 40\%;
the session completion signal removes the fundamental tradeoff of idle-time
reclamation 
for single-session state; and session-aware reuse classification identifies
shared state 4--700$\times$ more selectively than request-level access counts.
These results motivate a small cross-layer interface for workload-aware KV
storage without requiring application-specific semantics.
\end{abstract}

]

\printAffiliationsAndNotice{\textsuperscript{$\dagger$}Work done during an internship at IBM Research. }\

\section{Introduction}
\label{sec:intro}

Large language models (LLMs) now power applications ranging from interactive
chat and API services to coding assistants and autonomous
agents~\cite{wang2024burstgpt, patel2024splitwise, yao2023react,
yang2024sweagent}.
Serving these models efficiently depends heavily on the \emph{KV cache}: the
key and value tensors computed for previously processed
tokens~\cite{kwon2023vllm}. 
When requests share a prefix (such as a system prompt, retrieved document, or
conversation history), the corresponding KV state can be reused, avoiding
redundant prefill computation~\cite{kwon2023vllm, zheng2023sglang,
  gim2024promptcache}.

\begin{figure*}
    \centering
    \includegraphics[width=1\textwidth]{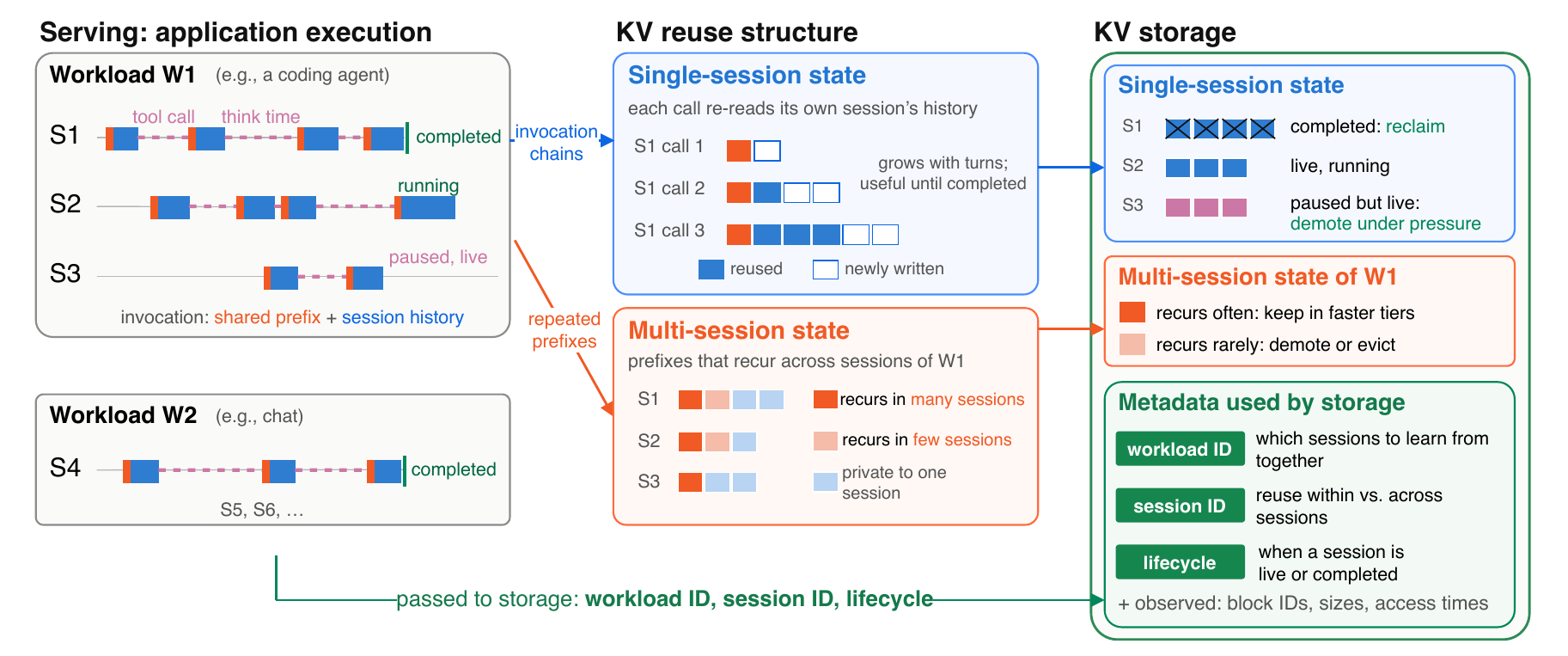}
    \caption{
\textbf{Execution structure underlying KV-cache reuse.}
Workloads contain multiple sessions, each comprising a sequence of model
invocations. This structure produces \emph{single-session state}, reused within
one session, and \emph{multi-session state}, reused across sessions.
KV storage can infer this structure from a small amount of additional metadata.
    }
    \label{fig:big-picture}
\end{figure*}

Long contexts, large KV footprints, and many concurrent sessions can quickly
exhaust GPU memory~\cite{kwon2023vllm, lin2024infinitellm,
sheng2023flexgen}.
Serving systems therefore increasingly extend KV caches into host DRAM, local
SSDs, and remote or distributed storage, retaining KV state so that it can be
reloaded rather than recomputed~\cite{gao2024cachedattention,
yu2025pensieve, qin2024mooncake, liu2025lmcache}.
Systems such as Mooncake~\cite{qin2024mooncake},
LMCache~\cite{liu2025lmcache}, SGLang HiCache~\cite{zheng2023sglang}, and
vLLM KV offloading~\cite{kwon2023vllm} 
make KV storage a first-class component of the serving stack.

We use \emph{KV storage} to refer to this multi-tier storage and management
layer: the component that retains KV blocks beyond the invocation that creates
them, places them across available tiers, and makes them available for reuse.
Its central decisions are therefore what state to retain, how long to retain
it, and where to place it.
Existing KV stores largely make these decisions using generic cache signals such
as recency and fixed time-to-live (TTL)~\cite{liu2025lmcache,
qin2024mooncake, zheng2023sglang, kwon2023vllm}.
These signals capture block accesses, but not the execution structure around
them; for example, which accesses belong to one session, which sessions belong
to the same recurring workload, or when a session has completed.

Prior systems demonstrate that exposing information from above the storage
layer can improve KV management.
They use application structure, workflow graphs, tool-call behavior, or
agent-specific information to guide retention, prefetching, and
eviction~\cite{lin2024parrot, li2025continuum,
tiwari2026cachewiseunderstandingworkloadsoptimizing, ren2025agserve}.
This raises a broader question for a shared KV store serving heterogeneous
applications: \emph{how much information from above does storage actually
need?}
Can a general-purpose store adapt to workload behavior using a \textbf{small,
  generic interface} rather than understanding each application's semantics?

Existing workload studies do not directly answer this question.
Request-level studies characterize arrivals and token
counts~\cite{wang2024burstgpt, patel2024splitwise};
workload-specific studies characterize sessions, tool calls, and agent steps
within individual application families
~\cite{zhu2026tracelabcharacterizingcodingagent,
yuan2026agenticaiworkloadcharacteristics}; and KV-cache studies examine reuse
within particular deployments~\cite{wang2025kvcachewild}.
What is missing is a \textbf{cross-workload view} of the KV blocks a shared
storage layer 
actually manages: where their reuse comes from, when and how frequently they
recur, how long they remain useful, and how their footprint evolves.

We characterize KV-cache reuse, lifetime, and footprint at the granularity of
KV-cache blocks across 11 production and public traces, which contain 15
workloads (Qwen-Bailian and $\tau^2$-bench each contribute three) in six
categories: chat, API serving, RAG, coding agents, tool-using agents, and
multi-lineage workflows.
For traces containing request contents, we reconstruct the prefix-chained
KV-block accesses visible to storage; we combine these with released block-level
traces and production traces containing timing and execution metadata.
Unless otherwise noted, we measure workload behavior without imposing a
particular cache capacity or eviction policy.

Figure~\ref{fig:big-picture} illustrates the execution structure that motivates
our analysis.
A workload is a recurring serving context containing multiple sessions,
with each session comprising a sequence of model invocations.
At the storage layer, this hierarchy induces two distinct scopes of reuse:
state reused within one session (\emph{single-session state}) and state that recurs across sessions of the
same workload (\emph{multi-session state}).
Our analysis reveals four main insights.

\stitle{Reuse scope is critical when deciding which KV state to retain.}
As sessions deepen, reuse increasingly comes from repeatedly reading the
session's own accumulated history, while shallow workloads can derive most of
their reuse from content recurring across sessions, such as system prompts,
templates, and retrieved documents.
Consequently, similar aggregate hit ratios can represent different storage
behavior: RAGBench and WildChat achieve similar potential hit ratios
(71\% and 74\%), but the former is dominated by multi-session reuse while the
latter primarily reuses single-session history.
Branching further weakens the relationship between session length and reuse:
Open Deep Research averages 39.5 model invocations (turns) per task but achieves only
20\% potential reuse because those invocations are spread across many short
context lineages.

The distinction matters because the two forms of reuse imply different
retention horizons and beneficiaries.
A hit on single-session state indicates value tied to one live execution, so
storage should manage that state with the session and reclaim it when the
session ends.
A hit on multi-session state demonstrates value beyond the current execution,
so storage should evaluate it based on recurrence across sessions rather than
the lifetime of any one session, using recurrence to decide whether it belongs
in a fast tier or a colder one.
Thus, aggregate hit rate alone is insufficient for retention and placement;
\emph{storage must know the scope of reuse}.

\stitle{Idle time is a poor signal for when KV state should be discarded.}
Single-session state can remain useful through long periods with no model
activity, including human think time and tool execution, but becomes dead when
the session completes.
Using an idle timer to approximate this boundary is difficult because the right
timeout varies substantially across workloads.
In Qwen To-C (a chat workload), a one-minute timeout retains only 22\% of within-session hits and even a
five-minute timeout loses 27\%.
Yet longer timeouts retain substantial dead state after sessions complete:
with a five-minute timeout, completed-session state accounts for 51\% of
resident token-seconds in Qwen To-C, 61\% in RAGPulse, and 40\% in
BrowseComp+.
No single timeout avoids both premature eviction and unnecessary retention
across these workloads.
An explicit session-completion signal therefore gives storage a cleaner
reclamation boundary for single-session state, while multi-session state must
instead be managed according to its recurrence across sessions.

\stitle{The source of capacity pressure determines the right storage response.}
Single-session state and multi-session state put pressure on cache capacity in
different ways. 
Single-session state grows as sessions deepen and accumulates across
concurrently live sessions.
Stalls can leave much of this state idle but still live: with 100 sessions
live at once, paused sessions account for 27--96\% of resident
single-session state across the production and agent traces we study.
Multi-session state creates a different problem.
Although it is stored only once across sessions, its shared pool contains a
large low-frequency tail: in seven of the 13 workloads with block identities, 71--91\% of
multi-session memory belongs to blocks accessed by at most five sessions.

These two sources of pressure call for different controls, regardless of whether
workloads share one KV store or are partitioned across separate storage pools. 
For single-session state, storage should track capacity by session, demote
paused-but-live state when pressure rises, and reclaim it when the session
completes. 
For multi-session state, storage should rank blocks by observed recurrence
within the relevant workload or storage partition, keeping frequently reused
blocks in faster tiers while demoting or evicting the low-frequency tail. 
The key distinction is therefore not whether different workload types compete
directly for the same capacity, but that single-session pressure is driven by
live-session footprint and concurrency, whereas multi-session pressure is driven
by how broadly and frequently shared state recurs. 
Treating both populations with the same retention policy can either evict state
still needed by live sessions or retain rarely reused shared state at the
expense of more valuable KV state. 

\stitle{A small amount of execution context exposes substantial management
  opportunities.} 
Storage already observes block identities, accesses, footprint, recurrence, and
capacity pressure.
Our findings suggest that much of the missing execution structure can be exposed
through three generic signals from the serving layer: an opaque workload
identifier, a session identifier, and session completion signal.
We directly quantify what each signal adds beyond block-level observations
alone.
Scoping idle-time behavior by workload retains up to 22\% more within-session
hits or reduces dead state by up to 40\% compared with a single global timeout.
Session completion eliminates the idle-time tradeoff for single-session state,
retaining useful state while a session is live and reclaiming it when the
session ends.
Finally, counting reuse across distinct sessions rather than across requests
identifies shared state 4--700$\times$ more selectively, avoiding repeated
private accesses being mistaken for cross-session sharing.

These measurements do not imply that the three signals are sufficient for an
optimal KV-management policy, nor do they predict an end-to-end throughput or
TTFT improvement.
Rather, they show that generic execution context resolves ambiguities that
materially change which KV state a storage policy would retain, reclaim, or
demote.
This provides a path toward workload-aware KV storage without requiring storage
to interpret whether an application is chat, RAG, an agent, or another workload
class.

In summary, this paper makes three contributions:
\begin{itemize}
    \item We characterize KV-cache reuse, lifetime, and footprint at KV-block
granularity across 11 production and public traces spanning six categories,
using a common methodology for heterogeneous traces.

\item We identify reuse scope as a key organizing principle: single-session
and multi-session state exhibit different reuse, lifetime, reclamation, and
capacity behavior.

\item We quantify the value of generic execution context: workload identity
scopes adaptation, session identity distinguishes private from shared reuse,
and lifecycle identifies reclaimable single-session state, without requiring
application-specific semantics.
\end{itemize}

\section{Background}
\label{sec:background}

\subsection{KV Blocks and Multi-Tier Storage}

\stitle{KV blocks and prefix reuse.}
During LLM inference, the serving engine (e.g., vLLM~\cite{kwon2023vllm},
SGLang~\cite{zheng2023sglang}) retains the key and value tensors 
computed for previously processed tokens so that later decoding steps and model
invocations need not recompute them.
KV-cache capacity grows with context length and model geometry and can consume a
large fraction of GPU memory~\cite{kwon2023vllm, sheng2023flexgen,
lin2024infinitellm}.
Modern serving engines therefore manage KV state in fixed-size
\emph{KV blocks}, each representing the state for a range of tokens across the
model's layers~\cite{kwon2023vllm, zheng2023sglang, liu2025lmcache}.
For example, vLLM's PagedAttention maps logical KV blocks to physical memory
through a block table~\cite{kwon2023vllm}.
KV blocks are also the granularity at which state is identified, reused,
transferred, and evicted, making them the natural unit for our analysis.

A cached block can be reused when a later invocation encounters the same token
sequence following the same prefix.
Prefix caching in these engines exploits this property to avoid
redundant prefill computation.
Reuse can occur within one execution as later turns revisit accumulated
history, or across executions through recurring system prompts, templates,
documents, or other shared prefixes~\cite{gim2024promptcache,
qin2024mooncake, wang2025kvcachewild}.
Useful KV state can therefore outlive the invocation that created it.

\stitle{Multi-tier KV storage.}
GPU memory cannot retain all potentially reusable KV state as contexts and
concurrency grow.
Systems therefore extend KV capacity into host DRAM, SSDs, and remote or
disaggregated memory~\cite{sheng2023flexgen, lee2024infinigen,
gao2024cachedattention, yu2025pensieve, chen2025impress}.
LMCache~\cite{liu2025lmcache}, Mooncake~\cite{qin2024mooncake},
SGLang HiCache~\cite{zheng2023sglang}, and related systems retain and move KV
blocks across such tiers.
Because the tiers differ in latency, bandwidth, and capacity, KV storage must
decide which blocks to retain, when to reclaim them, and where to place them.
Existing mechanisms primarily use storage-visible signals such as accesses,
recency, occupancy, and TTL, which describe block behavior but not the
execution structure that produced it.

\subsection{Execution Context for KV Management}

\stitle{What storage sees.}
A KV storage layer naturally observes the objects it manages: block identities
and sizes, reads and writes, access times, recurrence, occupancy, and pressure
on its storage tiers.
This information is sufficient to measure how often and how recently a block
has been accessed, but it does not identify the execution boundaries around
those accesses.

Three pieces of execution context are particularly relevant to our study.
A \emph{workload identifier} groups sessions belonging to the same recurring
application or serving context and gives storage a scope over which to learn
cross-session behavior.
A \emph{session identifier} groups accesses belonging to one logical execution
and lets storage distinguish reuse within a session from reuse across sessions.
Finally, \emph{session lifecycle} indicates when an execution has completed,
providing an explicit reclamation boundary for state whose value is confined
to that session.
These identifiers can be opaque to storage; they need not reveal application
semantics.

\begin{tcolorbox}[
    colback=gray!8,
    colframe=gray!35,
    boxrule=0.4pt,
    arc=0pt,
    left=5pt,
    right=5pt,
    top=5pt,
    bottom=5pt,
    before skip=6pt,
    after skip=6pt
]
\centering
\emph{Our study asks whether storage-visible block behavior, augmented with a
small amount of generic workload and session context, can capture the
distinctions that pertain to retention, reclamation, and placement.}
\end{tcolorbox}

\subsection{Related Work}
\label{sec:related}

\stitle{KV-cache storage and management.}
A growing body of work extends KV capacity beyond GPU memory and develops
mechanisms for retaining, transferring, and managing reusable KV state.
CachedAttention~\cite{gao2024cachedattention} and
Pensieve~\cite{yu2025pensieve} target stateful multi-turn serving, while
Mooncake~\cite{qin2024mooncake}, LMCache~\cite{liu2025lmcache}, and
SGLang~\cite{zheng2023sglang} provide mechanisms for storing and moving KV
state across memory and storage tiers.
Other systems specialize management further; for example,
IMPRESS~\cite{chen2025impress} uses importance information for multi-tier KV
storage, and CacheGen~\cite{liu2024cachegen} compresses KV state for transfer.
Our work is complementary to these systems.
Rather than proposing another KV-management mechanism, we characterize the
reuse, lifetime, and capacity demand presented to such a storage layer across
heterogeneous workloads, and ask what execution context is needed for making
its management decisions.

\stitle{Application- and workflow-aware KV management.}
Several recent systems exploit information above the KV-storage layer.
Parrot~\cite{lin2024parrot} exposes semantic variables and application
dependencies, KVFlow~\cite{pan2025kvflowefficientprefixcaching} exploits
multi-agent workflow structure, and PBKV~\cite{zheng2026pbkv} predicts future
invocations in dynamic agent workflows.
MORI~\cite{xia2026mori} exploits tool-call idle windows for
offloading, and CacheWise~\cite{tiwari2026cachewiseunderstandingworkloadsoptimizing}
uses characteristics of coding-agent workloads to guide KV management.
Continuum~\cite{li2025continuum} and AgServe~\cite{ren2025agserve} are
related in making sessions explicit: Continuum retains a session's
KV state across tool calls, and AgServe performs session-aware eviction.

These systems demonstrate that execution context can improve KV management,
but generally target a particular application family or exploit
application-, agent-, tool-, or workflow-specific structure.
Our question is different: \textbf{across heterogeneous workloads, how much of the
storage-relevant execution structure can be represented using generic signals?}
Our characterization identifies workload identity, session identity, and
session lifecycle as a small common interface that exposes the distinctions
behind the storage behaviors we observe.
We do not claim that richer information is unnecessary; it may improve
prediction before reuse or future execution is observed.

\stitle{LLM workload characterization.}
Prior studies characterize LLM serving from several complementary perspectives.
BurstGPT~\cite{wang2024burstgpt} and
Splitwise~\cite{patel2024splitwise} characterize request arrivals, token
distributions, and serving behavior, while
KVCache in the Wild~\cite{wang2025kvcachewild} compares KV-cache behavior
across several request types in production, including consumer-facing and API
workloads.
More recent work studies increasingly complex agentic applications.
TraceLab~\cite{zhu2026tracelabcharacterizingcodingagent} and
CacheWise~\cite{tiwari2026cachewiseunderstandingworkloadsoptimizing}
focus on coding agents, while
AgentSysBench~\cite{chang2026agentsysbench} compares ten agentic applications,
including RAG, coding, research, and interactive agents, from an end-to-end
serving perspective.
Nixon et al.~\cite{nixon2026yearllmserving} provide a complementary
longitudinal view of production LLM serving across many users and models.

Our study provides a different cross-workload view: we \textbf{normalize}
conventional 
and agentic workloads spanning chat, API serving, RAG, coding agents,
tool-using agents, and multi-lineage workflows into a common KV-block
representation.
This lets us compare reuse scope, lifetime, recurrence, and capacity behavior
at the granularity where a shared multi-tier KV store makes retention,
reclamation, and placement decisions.
Prior characterization has not combined this \textbf{breadth
  of 
workload structures} with a \textbf{common storage-level analysis of KV state}.

\section{Methodology and Workloads}
\label{sec:methodology}

We characterize KV behavior from the perspective of a shared, multi-tier
storage layer. We normalize each trace into a stream of KV-block creations and
accesses, together with timestamps and execution metadata when available.

\subsection{Representation and Reuse Scope}
\label{sec:block-populations}

A \emph{workload} is a recurring serving context containing multiple
\emph{sessions}; a session is one logical execution
containing one or more model invocations (turns).
Workload and session IDs can be opaque.

We classify reused KV state by its scope.
A block accessed only within one session is \emph{single-session state}, while
a block accessed by multiple sessions of the same workload is
\emph{multi-session state}.
Correspondingly, repeated accesses from the same session are
\emph{within-session hits}, and accesses from another session are
\emph{cross-session hits}.
Given workload and session IDs, storage can distinguish these hit
scopes online.

Block classification is retrospective: a block observed in one session may
later become multi-session state.
We therefore classify blocks using their complete observed trace, while the
storage implications we derive use only information available at each access.
Richer annotations such as semantic roles and tool types are used only to
explain behavior; workload categories such as chat or RAG are descriptive and
are not workload identifiers.

\subsection{Workloads}
\label{sec:workloads-collection}

We analyze 11 production and public traces, which contain 15 workloads in six
categories. Qwen-Bailian and $\tau^2$-bench each contribute three workloads,
which we analyze separately because they show distinct KV-access behavior.
Table~\ref{tab:traces} in Appendix~\ref{sec:appendix-traces} summarizes the
source, scale, and fields available for each workload.
Open-ended \textbf{chat} includes WildChat~\cite{zhao2024wildchat},
BurstGPT~\cite{wang2024burstgpt}, and Qwen-Bailian
To-C~\cite{wang2025kvcachewild}; \textbf{API serving} uses Qwen-Bailian
To-B~\cite{wang2025kvcachewild}; \textbf{RAG} includes
RAGBench~\cite{friel2024ragbench} and RAGPulse~\cite{wang2025ragpulse};
\textbf{coding agents} include Exgentic SWE-bench~\cite{bandel2026exgentic,
jimenez2024swebench}, Qwen-Bailian Code~\cite{wang2025kvcachewild}, and
TraceLab~\cite{zhu2026tracelabcharacterizingcodingagent}; and \textbf{tool-using
agents} include AppWorld~\cite{trivedi2024appworld}, BrowseComp+
~\cite{chen2025browsecompplus}, and the three $\tau^2$-bench domains
~\cite{yao2024taubench} from Exgentic~\cite{bandel2026exgentic}.

Our \textbf{multi-lineage} workload is Open Deep
Research - ODR~\cite{langchain_open_deep_research}, a LangGraph application in which
a supervisor delegates work to concurrent sub-researchers, creating multiple
independently growing context lineages within one session.
Because no public trace exposes the model inputs needed to reconstruct this
behavior, we collect a trace by running 50 English tasks from DeepResearch
Bench~\cite{du2025deepresearchbench}.
We serve Qwen3-8B~\cite{qwen3technicalreport} with
vLLM~\cite{kwon2023vllm} on one NVIDIA A100 GPU and record every model request
and response.
Appendix~\ref{sec:appendix-traces} gives further collection details.

\subsection{Trace Normalization}
\label{sec:workloads-method}

Traces differ in model, tokenizer, and block size, which change how requests
map to KV blocks and bytes; we normalize them to a common block-level
representation so that differences reflect workloads, not deployments.
Blocks generated during decode are stored and become reusable from the session's 
next call; outputs of a session's final call are never read and count only toward footprint.
Each analysis uses only workloads that provide the required fields.

\stitle{Raw-text traces.}
WildChat, RAGBench, the Exgentic workloads, and Open Deep Research provide
model inputs and session structure.
We tokenize requests using the Qwen3-8B
tokenizer~\cite{qwen3technicalreport}, divide them into 16-token blocks, and
compute vLLM-style prefix-chained hashes.
Blocks match only when they contain the same tokens following the same prefix.
Because the tokenizer may differ from that used in the original deployment,
these traces are normalized replays of request structure rather than exact
reconstructions of the original KV layout.
Appendix~\ref{sec:block-sensitivity} evaluates sensitivity to block size.

\stitle{Pre-hashed traces.}
Qwen-Bailian and RAGPulse provide block identifiers directly.
Qwen-Bailian uses 16-token prefix-chained blocks and includes session linkage.
RAGPulse identifies 512-token content chunks independently of prefix position,
so we chain its identifiers along each prompt to obtain prefix-exact identities.
Its larger blocks hide shorter shared prefixes, so we use RAGPulse primarily
for timing and lifetime analyses rather than direct reuse comparisons.

\stitle{Metadata-only traces.}
BurstGPT and the production timing portion of TraceLab provide timestamps,
token counts, and session structure but not reconstructable block identities.
We use these data only for idle-time, concurrency, and projected
footprint analyses; Appendix~\ref{sec:appendix-traces} describes how we use them in detail.

\section{Cross-Workload Analysis}
\label{sec:analysis}

We examine three properties of KV state that affect multi-tier storage:
reuse scope (\S\ref{sec:reuse}), lifetime (\S\ref{sec:temporal}), and capacity
demand (\S\ref{sec:volume}).
Unless noted otherwise, measurements assume unlimited capacity and no eviction,
so they characterize behavior in the access stream rather than behavior induced
by a particular cache policy.

\subsection{Reuse Scope and Concentration}
\label{sec:reuse}

\finding{1}{Reuse scope determines how storage should manage reused KV state.}

\begin{figure}
    \centering
    \includegraphics[width=1\linewidth]{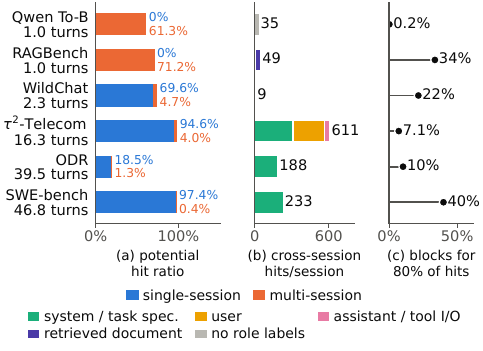}
    \caption{Reuse in six representative workloads, ordered by turns per
    session. (a) Potential hits split into within-session and cross-session
    reuse. (b) Cross-session hits per session, broken down by semantic role
    where available. (c) Fraction of unique blocks supplying 80\% of hits.}
    \label{fig:reuse-sources}
\end{figure}

Substantial KV reuse occurs across workload types, but its source differs
sharply.
We define \emph{potential hit ratio} as the fraction of block reads that would
find the block already present under unlimited capacity.
Among the representative workloads, it ranges from 20\% in Open Deep Research (ODR)
to 99\% in $\tau^2$-Telecom (Figure~\ref{fig:reuse-sources}a).
Reuse is also concentrated: at most 40\% of unique blocks supply 80\% of hits
in every workload shown, and Qwen To-B requires only 0.2\%
(Figure~\ref{fig:reuse-sources}c).

\subfinding{1.1}{Session structure determines the source of reuse; aggregate
hit ratios hide this distinction.}

Each turn in a linear session rereads context accumulated by earlier turns, so
deeper sessions naturally create more within-session reuse.
Single-turn Qwen To-B and RAGBench obtain all hits across session boundaries,
whereas WildChat, with 2.3 turns per session, already obtains most reuse from
its own history.
$\tau^2$-Telecom and SWE-bench, with mean 16.3 and 46.8 turns per session,
respectively, reach 99\% and 98\% potential hit ratios dominated by
within-session reuse.

Similar aggregate hit ratios can therefore represent different storage
behavior.
RAGBench and WildChat achieve 71\% and 74\% potential hit ratios, but RAGBench
is dominated by cross-session reuse while WildChat primarily rereads
single-session history.

Branching breaks a simple relationship between task length and reuse.
ODR averages 39.5 model invocations per task but only 20\%
potential reuse because those invocations are distributed across roughly 33
independently evolving context lineages, each averaging 1.2 turns.
The relevant quantity is therefore the depth of a prefix lineage rather than
the total number of invocations in the task.

Cross-session fractions can also hide substantial absolute reuse
(Figure~\ref{fig:reuse-sources}b).
Only 0.4\% of SWE-bench hits cross sessions, yet each session has 233
cross-session hits, compared with 9 in WildChat;
99\% are on the system prompt and tool definitions shared across independent coding problems.
$\tau^2$-Telecom has 611 cross-session hits per session.
Deep workloads can therefore contain substantial single- and multi-session
reuse simultaneously, even when the latter is a small fraction of all hits.

\storageimplication{
Reuse scope should guide retention and placement.
Single-session reuse is valuable only while its session remains live, whereas
cross-session reuse provides evidence of value beyond the current execution.
Aggregate hit rate alone therefore does not reveal how long or where reused
state should be retained.
}

\subfinding{1.2}{Prefix recurrence, not semantic role, determines sharing.}

Cross-session reuse requires identical content to recur after an identical
prefix.
Stable system prompts often satisfy this condition, but other semantic regions
can as well.
Using trace annotations, we group blocks into system or task-specification,
user, assistant or tool-I/O, and retrieved-document regions.
In $\tau^2$-Telecom, user blocks contribute roughly two-fifths of its 611
cross-session hits per session, with assistant and tool blocks contributing
substantial additional reuse (Figure~\ref{fig:reuse-sources}b).

Loaded content illustrates the same principle.
RAGBench repeatedly draws from a fixed document corpus and obtains 46 of
its 49 cross-session hits per session from retrieved content, whereas ODR issues diverse web queries and obtains essentially no cross-session
reuse from retrieved pages.
Large retrieved inputs therefore do not imply reusable KV state: content must
recur under a stable prefix. ODR also places the same research brief after
different prefixes in different lineages, preventing reuse.

\storageimplication{
Storage need not understand semantic roles to recognize realized sharing.
Observed recurrence across sessions provides a generic signal for identifying
and ranking multi-session state; semantic hints may still help predict reuse
before it occurs.
}

\subsection{Lifetime and Recurrence}
\label{sec:temporal}

\finding{2}{Idle time is a poor signal for when KV state should be discarded.}

Single-session state can remain useful while its session is idle, but its value
ends when that session completes.
Without lifecycle information, storage must approximate this boundary using an
idle timeout, which can fail in both directions.

\subfinding{2.1}{Setting time to live is challenging for both single-session and
  multi-session state.}

\begin{figure}
    \centering
    \includegraphics[width=1\linewidth]{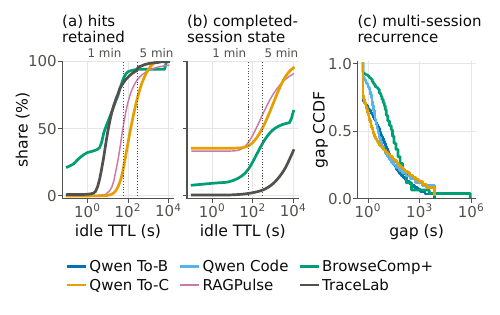}
    \caption{Lifetime of single- and multi-session state. Using an idle timer to approximate single-session lifetime: 
    (a) within-session hits retained and (b) share of resident token-seconds
    held by sessions that have already completed, as the timer grows;
    (c) recurrence gaps for multi-session state; tails are bounded by collection window (Table~\ref{tab:traces}).}
    \label{fig:temporal-ttl}
\end{figure}

The appropriate timeout varies substantially across workloads
(Figure~\ref{fig:temporal-ttl}a, b).
In Qwen To-C, a one-minute timer retains only 22\% of within-session hits and a
five-minute timer still retains only 73\%.
BrowseComp+ retains 88\% with a one-minute timer, while TraceLab retains 48\% with 
a ten-second timer.
Longer timers reduce premature expiration but retain dead state: with a
five-minute timer, completed-session state accounts for 51\% of resident
token-seconds in Qwen To-C, 61\% in RAGPulse, and 40\% in BrowseComp+,
compared with 4\% in TraceLab.
No single timeout avoids both errors across these workloads. 

When available, session completion removes this ambiguity for single-session
state by providing an explicit reclamation boundary.
Some workloads, particularly open-ended chat, may not expose a definitive
completion event.
In such cases, idle time can still guide demotion to a lower tier, but cannot
reliably distinguish a paused session from one whose state is safe to reclaim.

Multi-session state has no session-completion boundary because completing one
session does not invalidate state that may recur in another.
Instead, its recurrence gaps span several orders of magnitude
(Figure~\ref{fig:temporal-ttl}c).
In the four workloads shown, 69--80\% of gaps are under a
minute, yet 11--18\% exceed five minutes and 4--10\% exceed an hour.
Counted in intervening sessions, the spread is wider still: BrowseComp+'s
common prefix recurs almost every session, whereas half of Qwen To-B's gaps
exceed 100 sessions (Appendix~\ref{sec:appendix-recurrence}).
Thus, multi-session state ranges from continuously recurring prefixes to blocks
unused for hours.

\storageimplication{
The two reuse scopes need different lifetime signals.
Session completion, when available, gives single-session state an explicit
reclamation boundary; otherwise idle time can guide demotion but not definitive
reclamation.
Multi-session state should instead be ranked by recurrence across sessions.
}

\subfinding{2.2}{Stalls create idle-but-live state whose cost grows with
overlapping demand.}

Tool execution, API latency, and human think time can leave a live session
inactive even though later invocations still require its KV state.
This paused-but-live state consumes fast-tier capacity, but reclaiming it
would force a later reload or recomputation; the pressure grows as more live
sessions overlap.

\begin{figure}
    \centering
    \includegraphics[width=1\linewidth]{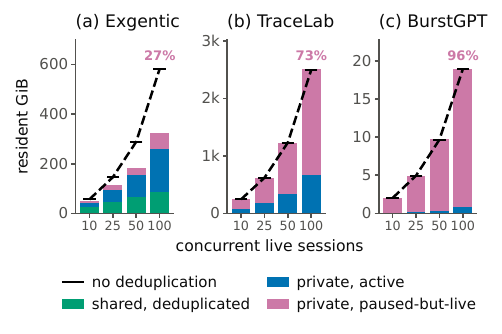}
    \caption{Resident KV state under increasing concurrency, split into
    multi-session state and active versus paused single-session state.
    Percentages report the paused share of single-session state at
    100 live sessions. Exgentic pools its six agent workloads.}
    \label{fig:temporal-concurrency}
\end{figure}

With 100 sessions live at once, paused sessions account for 27\% of
single-session resident state in Exgentic, 73\% in TraceLab, and 96\% in
BurstGPT (Figure~\ref{fig:temporal-concurrency}; setup in Appendix~\ref{sec:appendix-temporal}).
The absolute footprint depends on how much state each session has accumulated:
TraceLab's long coding sessions create substantially more resident KV state than
the short conversational sessions in BurstGPT at the same concurrency.

\storageimplication{
Inactivity should not be treated as death.
Under capacity pressure, storage can demote paused-but-live state while session
lifecycle distinguishes it from state that can be reclaimed.
}

\subsection{Footprint and Capacity}
\label{sec:volume}

\finding{3}{The source of capacity pressure determines the right storage
response.}

\begin{figure*}[t]
    \centering
    \includegraphics[width=\textwidth]{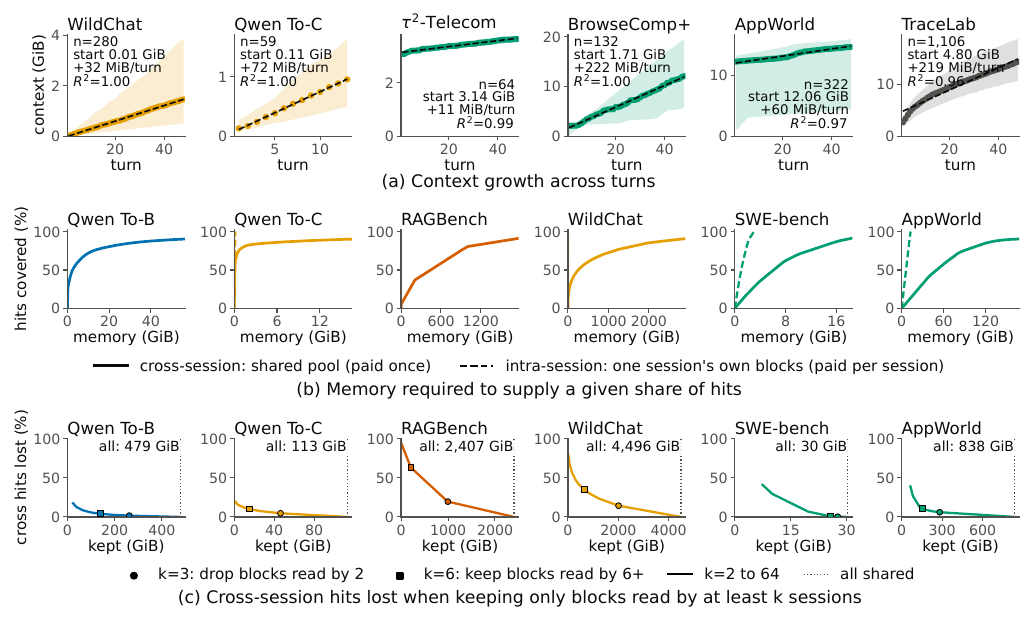}
    \caption{KV footprint across representative workloads.
    (a) Context growth with session depth.
    (b) Memory needed to supply a given fraction of cross-session hits.
    (c) Multi-session memory retained and hits lost when filtering blocks by
    the number of sessions that reuse them.}
    \label{fig:volume}
\end{figure*}

Single-session and multi-session state consume capacity differently.
Single-session state grows with each live execution and is replicated across
concurrent sessions.
Multi-session state is stored once across sessions, but can accumulate a large
tail of blocks that recur infrequently.
For a common byte scale, we normalize tokens using the KV geometry of Qwen3-8B
in \texttt{bf16} (144\,KiB per token).
Appendix~\ref{sec:appendix-volume} reports the corresponding measurements for
all workloads.

\subfinding{3.1}{Session footprint depends on both initial context and growth.}

Figure~\ref{fig:volume}a shows how single-session footprint evolves with
session depth.
Across multi-turn workloads, median context grows approximately linearly with
session depth ($R^2 \ge 0.95$), but both starting footprint, determined by
initial context such as system prompts, tool definitions, and loaded content,
and subsequent growth rate vary widely.
AppWorld begins at 12.1\,GiB and $\tau^2$-Telecom at 3.1\,GiB, whereas
WildChat and Qwen To-C begin at 0.01 and 0.11\,GiB.
Per-turn growth varies by roughly $20\times$, from 11\,MiB in
$\tau^2$-Telecom to 222\,MiB in BrowseComp+.
These two quantities can reverse which workload dominates capacity:
BrowseComp+ begins at roughly half the footprint of $\tau^2$-Telecom but,
because it grows about $20\times$ faster per turn, surpasses it within ten
turns.

Session depth determines how far this growth accumulates.
The median $\tau^2$-Telecom session lasts 13 turns but its 99th percentile
reaches 105; TraceLab has a median of 17 turns but a 99th percentile above
1,000.
Neither initial context size nor typical session length alone therefore
captures capacity demand.

\storageimplication{
Single-session capacity should be accounted for per live session rather than
through an application-wide fixed reservation.
Storage can estimate demand from each session's observed footprint and growth,
combined with the number of concurrently live sessions.
}

\subfinding{3.2}{Most multi-session KV volume belongs to blocks reused by few
sessions.}

Figure~\ref{fig:volume}b compares how much memory is required to capture
within-session and cross-session reuse.
The private state of only six concurrent SWE-bench sessions already exceeds
the 17.6\,GiB required to supply 90\% of its cross-session hits.

The multi-session pool contains a pronounced low-frequency tail.
Frequently recurring prefixes form a compact hot head: half of cross-session
hits require only 0.09\,GiB in Qwen To-C and 2.4\,GiB in Qwen To-B.
The complete Qwen To-B multi-session pool, however, reaches 479\,GiB.
RAGBench is less concentrated, requiring 443\,GiB to supply half of its
cross-session hits.

Figure~\ref{fig:volume}c shows that much of the multi-session pool is reused
by only a few sessions.
In five of the six workloads shown, blocks accessed by at most five sessions
account for 71--91\% of multi-session memory; Appendix~\ref{sec:appendix-volume}
reports all workloads.
Their usefulness nevertheless differs across workloads: discarding these blocks
loses only 4\% of cross-session hits in Qwen To-B but 63\% in RAGBench.
A fixed reuse-frequency threshold is therefore unlikely to work across
workloads.

\storageimplication{
Multi-session state should not be treated as uniformly valuable.
Observed cross-session recurrence can rank the shared pool, keeping frequently
recurring blocks in faster tiers while demoting or evicting the low-frequency
tail.
}

\subsection{Application-Independent Execution Context}
\label{sec:analysis-interface}

Across these findings, \textbf{storage decisions depend on reuse scope,
  lifecycle, 
recurrence, and capacity cost rather than application type}.
Most of this information is already visible to storage: block accesses reveal
reuse and recurrence, allocations reveal footprint, and occupancy exposes
capacity pressure.
The missing information is execution structure: which sessions belong to the
same workload, which accesses belong to one session, and when that session
completes.

\begin{figure*}[t]
    \centering
    \includegraphics[width=1\linewidth]{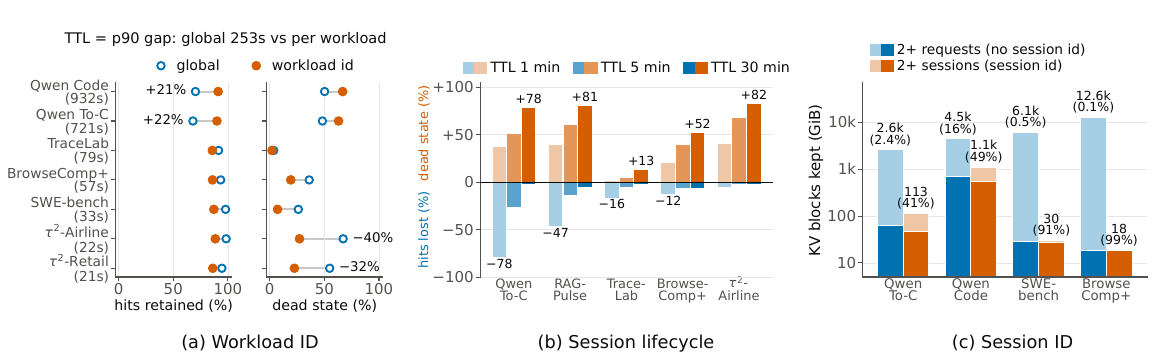}
    \caption{What each piece of execution context adds.
    (a) Workload ID: idle TTL set to the 90th percentile of observed
    within-session gaps, estimated over all seven workloads or over each workload's own sessions.
    (b) Session lifecycle: idle TTL of 1, 5, and 30 minutes compared with
    reclaiming state at session completion.
    (c) Session ID: blocks flagged as reused at their second access, counted per
    request (no session ID) or per distinct session (session ID), with causal
    counts and no capacity limit. Each full bar is the state kept as reused;
    its dark part is the state that a later session actually reads, with that
    share in parentheses.}
    \label{fig:info}
\end{figure*}

Figure~\ref{fig:info} quantifies what each of the three execution signals
adds beyond block-level observations alone.

\stitle{Workload ID.}
Storage can learn a timeout value across all the sessions it serves;
however, with workload identity, storage can adapt the timeout value to each
workload to be significantly more accurate. 
Figure~\ref{fig:info}a illustrates this difference by
setting the TTL 
to the 90th percentile of observed reuse.
The pooled TTL of 253\,s is too short for Qwen To-C and Qwen Code:
per-workload estimates retain 22\% and 21\% more within-session hits.
Conversely, per-workload estimates reduce dead state (state that can be safely
dropped but is retained) by 40\% in 
$\tau^2$-Airline and 32\% in $\tau^2$-Retail.
The workload ID does not improve both metrics for every workload; it lets each
workload operate a timeout suited to its own reuse distribution.

\stitle{Session lifecycle.}
Figure~\ref{fig:info}b compares idle timeouts with reclamation at
session completion.
Completion provides an explicit end-of-life signal for 
single-session state: it preserves all within-session hits while immediately
reclaiming state that can no longer be reused. 
No fixed timeout achieves both.
A one-minute TTL loses 78\% of within-session hits in Qwen To-C and 47\% in
RAGPulse; a 30-minute TTL keeps nearly all of them, but dead state then makes
up 78--82\% of held state in Qwen To-C, RAGPulse, and $\tau^2$-Airline, 
and 52\% in BrowseComp+.
The completion signal removes this trade-off for single-session state.

Completion alone is insufficient for handling multi-session state: in Qwen Code,
where 
53\% of re-reads cross sessions, reclaiming every block no longer held by a
live session retains 37\% fewer hits than a five-minute TTL
(Appendix~\ref{sec:appendix-completion}).
Lifecycle-based reclamation therefore requires session identity to distinguish
single-session from shared state.

\stitle{Session ID.}
Without session identity, repeated reads within one execution can be mistaken
for sharing across executions.
Figure~\ref{fig:info}c flags a block at its second access, counted either per
request or per distinct session.

Counting requests flags 2.6k--12.6k\,GiB in the four workloads, of which only
0.1--16\% is later read by another session.
Using session IDs flags 4--700$\times$ less state (18\,GiB--1.1k\,GiB), and
41--99\% of it is read by a later session.
The session ID thus identifies the shared state that completion should spare,
so lifecycle can reclaim single-session state without discarding blocks that
later sessions read (Appendix~\ref{sec:appendix-completion}).

Together, the three signals resolve different ambiguities: workload ID
scopes adaptation, session ID distinguishes private reuse from sharing,
and session completion exposes liveness.
Figure~\ref{fig:info} shows
that each materially changes which state a policy would retain or reclaim
without requiring application-specific semantics.

\section{Discussion}
\label{sec:discussion}

\stitle{Scope of the characterization.}
Our trace-driven analysis characterizes demand presented to KV storage rather
than a complete serving stack, and does not model transfer costs, scheduling,
or a particular finite-capacity policy.
Some statistics are measured offline rather than learned incrementally.
Our results quantify the information exposed by workload, session, and
lifecycle context, but do not predict the end-to-end performance of a specific
storage policy.

\stitle{Execution boundaries and trace normalization.}
We assume the serving layer can associate requests with workloads and sessions
and, when available, expose session completion.
Raw-text traces are normalized using one tokenizer, 16-token blocks, and the
KV geometry of Qwen3-8B; Appendix~\ref{sec:block-sensitivity} shows that the
main reuse and recurrence trends are stable across block sizes.

\stitle{Complementary techniques}.
We study prefix-exact reuse within each workload.
Reuse across differing prefixes~\cite{yao2024cacheblend}, attention-based KV
selection~\cite{zhang2023h2o, li2024snapkv}, and cross-workload reuse are
complementary to our work.

\section{Conclusion}
\label{sec:conclusion}

We present a storage-oriented characterization of KV-cache behavior across
heterogeneous LLM workloads.
We show that three pieces of generic execution context can materially
improve how KV storage manages state.
These results motivate workload-aware KV storage built on a small
cross-layer interface rather than application-specific semantics.


\bibliographystyle{mlsys2025}
\bibliography{reference}

@inproceedings{gao2024cachedattention,
      title={Cost-Efficient Large Language Model Serving for Multi-turn Conversations with {CachedAttention}},
      author={Bin Gao and Zhuomin He and Puru Sharma and Qingxuan Kang and Djordje Jevdjic and Junbo Deng and Xingkun Yang and Zhou Yu and Pengfei Zuo},
      year={2024},
      booktitle={Proceedings of the 2024 USENIX Annual Technical Conference (ATC)},
      url={https://www.usenix.org/conference/atc24/presentation/gao-bin-cost},
}

@inproceedings{yu2025pensieve,
      title={Stateful Large Language Model Serving with {Pensieve}},
      author={Lingfan Yu and Jinkun Lin and Jinyang Li},
      year={2025},
      booktitle={Proceedings of the Twentieth European Conference on Computer Systems (EuroSys)},
      doi={10.1145/3689031.3696086},
}

@inproceedings{gim2024promptcache,
      title={Prompt Cache: Modular Attention Reuse for Low-Latency Inference},
      author={In Gim and Guojun Chen and Seung-seob Lee and Nikhil Sarda and Anurag Khandelwal and Lin Zhong},
      year={2024},
      booktitle={Proceedings of Machine Learning and Systems (MLSys)},
}

@inproceedings{yao2023react,
      title={{ReAct}: Synergizing Reasoning and Acting in Language Models},
      author={Shunyu Yao and Jeffrey Zhao and Dian Yu and Nan Du and Izhak Shafran and Karthik Narasimhan and Yuan Cao},
      year={2023},
      booktitle={International Conference on Learning Representations (ICLR)},
}

@software{langchain_open_deep_research,
  author       = {{LangChain AI}},
  organization = {LangChain AI},
  title        = {Open Deep Research},
  year         = {2025},
  url          = {https://github.com/langchain-ai/open_deep_research},
  note         = {GitHub repository}
}

@misc{xia2026mori,
      title={Idleness is Relative: Exploiting Tool-Call Idle Windows for Offloading in Agentic Systems with MORI},
      author={Tian Xia and Hanchen Li and Zhifei Li and Xiaokun Chen and Hao Kang and Yifan Qiao and Yi Xu and Ion Stoica},
      year={2026},
      eprint={2606.00866},
      archivePrefix={arXiv},
      primaryClass={cs.OS},
      url={https://arxiv.org/abs/2606.00866},
}

@misc{du2025deepresearchbench,
      title={DeepResearch Bench: A Comprehensive Benchmark for Deep Research Agents},
      author={Mingxuan Du and Benfeng Xu and Chiwei Zhu and Xiaorui Wang and Zhendong Mao},
      year={2025},
      eprint={2506.11763},
      archivePrefix={arXiv},
      primaryClass={cs.CL},
      url={https://arxiv.org/abs/2506.11763},
}

@misc{qwen3technicalreport,
      title={Qwen3 Technical Report},
      author={{Qwen Team}},
      year={2025},
      eprint={2505.09388},
      archivePrefix={arXiv},
      primaryClass={cs.CL},
      url={https://arxiv.org/abs/2505.09388},
}

@misc{li2025continuum,
      title={Continuum: Efficient and Robust Multi-Turn LLM Agent Scheduling with KV Cache Time-to-Live},
      author={Hanchen Li and Runyuan He and Qiuyang Mang and Qizheng Zhang and Huanzhi Mao and Xiaokun Chen and Hangrui Zhou and Alvin Cheung and Joseph Gonzalez and Ion Stoica},
      year={2025},
      eprint={2511.02230},
      archivePrefix={arXiv},
      primaryClass={cs.OS},
      url={https://arxiv.org/abs/2511.02230},
}

@misc{bandel2026exgentic,
      title={General Agent Evaluation},
      author={Elron Bandel and Asaf Yehudai and Lilach Eden and Yehoshua Sagron and Yotam Perlitz and Elad Venezian and Natalia Razinkov and Natan Ergas and Shlomit Shachor Ifergan and Segev Shlomov and Michal Jacovi and Leshem Choshen and Liat Ein-Dor and Yoav Katz and Michal Shmueli-Scheuer},
      year={2026},
      eprint={2602.22953},
      archivePrefix={arXiv},
      primaryClass={cs.AI},
      url={https://arxiv.org/abs/2602.22953},
}

@misc{yang2024sweagent,
      title={SWE-agent: Agent-Computer Interfaces Enable Automated Software Engineering},
      author={John Yang and Carlos E. Jimenez and Alexander Wettig and Kilian Lieret and Shunyu Yao and Karthik Narasimhan and Ofir Press},
      year={2024},
      eprint={2405.15793},
      archivePrefix={arXiv},
      primaryClass={cs.SE},
      url={https://arxiv.org/abs/2405.15793},
}

@misc{jimenez2024swebench,
      title={SWE-bench: Can Language Models Resolve Real-World GitHub Issues?},
      author={Carlos E. Jimenez and John Yang and Alexander Wettig and Shunyu Yao and Kexin Pei and Ofir Press and Karthik Narasimhan},
      year={2024},
      eprint={2310.06770},
      archivePrefix={arXiv},
      primaryClass={cs.CL},
      url={https://arxiv.org/abs/2310.06770},
}

@misc{wang2025ragpulse,
      title={RAGPulse: An Open-Source RAG Workload Trace to Optimize RAG Serving Systems},
      author={Zhengchao Wang and Yitao Hu and Jianing Ye and Zhuxuan Chang and Jiazheng Yu and Youpeng Deng and Keqiu Li},
      year={2025},
      eprint={2511.12979},
      archivePrefix={arXiv},
      primaryClass={cs.DC},
      url={https://arxiv.org/abs/2511.12979},
}

@misc{friel2024ragbench,
      title={RAGBench: Explainable Benchmark for Retrieval-Augmented Generation Systems},
      author={Robert Friel and Masha Belyi and Atindriyo Sanyal},
      year={2024},
      eprint={2407.11005},
      archivePrefix={arXiv},
      primaryClass={cs.CL},
      url={https://arxiv.org/abs/2407.11005},
}

@inproceedings{wang2025kvcachewild,
      title={KVCache Cache in the Wild: Characterizing and Optimizing KVCache Cache at a Large Cloud Provider},
      author={Jiahao Wang and Jinbo Han and Xingda Wei and Sijie Shen and Dingyan Zhang and Chenguang Fang and Rong Chen and Wenyuan Yu and Haibo Chen},
      year={2025},
      booktitle={Proceedings of the 2025 USENIX Annual Technical Conference (ATC)},
}

@misc{yao2024taubench,
      title={tau-bench: A Benchmark for Tool-Agent-User Interaction in Real-World Domains},
      author={Shunyu Yao and Noah Shinn and Pedram Razavi and Karthik Narasimhan},
      year={2024},
      eprint={2406.12045},
      archivePrefix={arXiv},
      primaryClass={cs.AI},
      url={https://arxiv.org/abs/2406.12045},
}

@misc{trivedi2024appworld,
      title={AppWorld: A Controllable World of Apps and People for Benchmarking Interactive Coding Agents},
      author={Harsh Trivedi and Tushar Khot and Mareike Hartmann and Ruskin Manku and Vinty Dong and Edward Li and Shashank Gupta and Ashish Sabharwal and Niranjan Balasubramanian},
      year={2024},
      eprint={2407.18901},
      archivePrefix={arXiv},
      primaryClass={cs.CL},
      url={https://arxiv.org/abs/2407.18901},
}

@misc{yuan2026agenticaiworkloadcharacteristics,
      title={Agentic AI Workload Characteristics},
      author={Yichao Yuan and Ankita Nayak and Souvik Kundu and Nishil Talati},
      year={2026},
      eprint={2605.26297},
      archivePrefix={arXiv},
      primaryClass={cs.DC},
      url={https://arxiv.org/abs/2605.26297},
}

@misc{zhu2026tracelabcharacterizingcodingagent,
      title={TraceLab: Characterizing Coding Agent Workloads for LLM Serving},
      author={Kan Zhu and Mathew Jacob and Chenxi Ma and Yi Pan and Stephanie Wang and Arvind Krishnamurthy and Baris Kasikci},
      year={2026},
      eprint={2606.30560},
      archivePrefix={arXiv},
      primaryClass={cs.LG},
      url={https://arxiv.org/abs/2606.30560},
}

@misc{wang2024burstgpt,
      title={BurstGPT: A Real-world Workload Dataset to Optimize LLM Serving Systems},
      author={Yuxin Wang and Yuhan Chen and Zeyu Li and Xueze Kang and Yuchu Fang and Yeju Zhou and Yang Zheng and Zhenheng Tang and Xin He and Rui Guo and Xin Wang and Qiang Wang and Amelie Chi Zhou and Xiaowen Chu},
      year={2024},
      eprint={2401.17644},
      archivePrefix={arXiv},
      primaryClass={cs.DC},
      url={https://arxiv.org/abs/2401.17644},
}

@inproceedings{patel2024splitwise,
      title={Splitwise: Efficient Generative LLM Inference Using Phase Splitting},
      author={Pratyush Patel and Esha Choukse and Chaojie Zhang and Aashaka Shah and Inigo Goiri and Saeed Maleki and Ricardo Bianchini},
      year={2024},
      booktitle={Proceedings of the 51st International Symposium on Computer Architecture (ISCA)},
      eprint={2311.18677},
      archivePrefix={arXiv},
      url={https://arxiv.org/abs/2311.18677},
}

@misc{zhao2024wildchat,
      title={WildChat: 1M ChatGPT Interaction Logs in the Wild},
      author={Wenting Zhao and Xiang Ren and Jack Hessel and Claire Cardie and Yejin Choi and Yuntian Deng},
      year={2024},
      eprint={2405.01470},
      archivePrefix={arXiv},
      primaryClass={cs.CL},
      url={https://arxiv.org/abs/2405.01470},
}

@misc{pan2025kvflowefficientprefixcaching,
      title={KVFlow: Efficient Prefix Caching for Accelerating LLM-Based Multi-Agent Workflows},
      author={Zaifeng Pan and Ajjkumar Patel and Zhengding Hu and Yipeng Shen and Yue Guan and Wan-Lu Li and Lianhui Qin and Yida Wang and Yufei Ding},
      year={2025},
      eprint={2507.07400},
      archivePrefix={arXiv},
      primaryClass={cs.DC},
      url={https://arxiv.org/abs/2507.07400},
}

@misc{tiwari2026cachewiseunderstandingworkloadsoptimizing,
      title={CacheWise: Understanding Workloads and Optimizing KVCache Management for Efficiently Serving LLM Coding Agents},
      author={Shubham Tiwari and Tapan Chugh and Nash Rickert and Simon Peter and Ratul Mahajan and Haiying Shen},
      year={2026},
      eprint={2606.16824},
      archivePrefix={arXiv},
      primaryClass={cs.DC},
      url={https://arxiv.org/abs/2606.16824},
}

@inproceedings{kwon2023vllm,
      author = {Kwon, Woosuk and Li, Zhuohan and Zhuang, Siyuan and Sheng, Ying and Zheng, Lianmin and Yu, Cody Hao and Gonzalez, Joseph and Zhang, Hao and Stoica, Ion},
      title = {Efficient Memory Management for Large Language Model Serving with PagedAttention},
      year = {2023},
      publisher = {Association for Computing Machinery},
      doi = {10.1145/3600006.3613165},
      booktitle = {Proceedings of the 29th Symposium on Operating Systems Principles},
      pages = {611--626},
      series = {SOSP '23}
}

@inproceedings{qin2024mooncake,
      title={Mooncake: Trading More Storage for Less Computation --- A {KVCache}-centric Architecture for Serving {LLM} Chatbot},
      author={Ruoyu Qin and Zheming Li and Weiran He and Jialei Cui and Feng Ren and Mingxing Zhang and Yongwei Wu and Weimin Zheng and Xinran Xu},
      year={2025},
      booktitle={Proceedings of the 23rd USENIX Conference on File and Storage Technologies (FAST)},
      url={https://www.usenix.org/conference/fast25/presentation/qin},
}

@misc{liu2025lmcache,
      title={LMCache: An Efficient KV Cache Layer for Enterprise-Scale LLM Inference},
      author={Yuhan Liu and Yihua Cheng and Jiayi Yao and Yuwei An and Xiaokun Chen and Shaoting Feng and Yuyang Huang and Samuel Shen and Rui Zhang and Kuntai Du and Junchen Jiang},
      year={2025},
      eprint={2510.09665},
      archivePrefix={arXiv},
      primaryClass={cs.DC},
      url={https://arxiv.org/abs/2510.09665},
}

@misc{zheng2023sglang,
      title={SGLang: Efficient Execution of Structured Language Model Programs},
      author={Lianmin Zheng and Liangsheng Yin and Zhiqiang Xie and Chuyue Sun and Jeff Huang and Cody Hao Yu and Shiyi Cao and Christos Kozyrakis and Ion Stoica and Joseph E. Gonzalez and Clark Barrett and Ying Sheng},
      year={2023},
      eprint={2312.07104},
      archivePrefix={arXiv},
      primaryClass={cs.CL},
      url={https://arxiv.org/abs/2312.07104},
}

@misc{liu2024cachegen,
      title={CacheGen: KV Cache Compression and Streaming for Fast Large Language Model Serving},
      author={Yuhan Liu and Hanchen Li and Yihua Cheng and Siddhant Ray and Yuyang Huang and Qizheng Zhang and Kuntai Du and Jiayi Yao and Shan Lu and Ganesh Ananthanarayanan and Michael Maire and Henry Hoffmann and Ari Holtzman and Junchen Jiang},
      year={2024},
      eprint={2310.07240},
      archivePrefix={arXiv},
      primaryClass={cs.LG},
      url={https://arxiv.org/abs/2310.07240},
}

@misc{yao2024cacheblend,
      title={CacheBlend: Fast Large Language Model Serving for RAG with Cached Knowledge Fusion},
      author={Jiayi Yao and Hanchen Li and Yuhan Liu and Siddhant Ray and Yihua Cheng and Qizheng Zhang and Kuntai Du and Shan Lu and Junchen Jiang},
      year={2024},
      eprint={2405.16444},
      archivePrefix={arXiv},
      primaryClass={cs.CL},
      url={https://arxiv.org/abs/2405.16444},
}

@misc{zhang2023h2o,
      title={H2O: Heavy-Hitter Oracle for Efficient Generative Inference of Large Language Models},
      author={Zhenyu Zhang and Ying Sheng and Tianyi Zhou and Tianlong Chen and Lianmin Zheng and Ruisi Cai and Zhao Song and Yuandong Tian and Christopher Re and Clark Barrett and Zhangyang Wang and Beidi Chen},
      year={2023},
      eprint={2306.14048},
      archivePrefix={arXiv},
      primaryClass={cs.LG},
      url={https://arxiv.org/abs/2306.14048},
}

@misc{li2024snapkv,
      title={SnapKV: LLM Knows What You are Looking for Before Generation},
      author={Yuhong Li and Yingbing Huang and Bowen Yang and Bharat Venkitesh and Acyr Locatelli and Hanchen Ye and Tianle Cai and Patrick Lewis and Deming Chen},
      year={2024},
      eprint={2404.14469},
      archivePrefix={arXiv},
      primaryClass={cs.CL},
      url={https://arxiv.org/abs/2404.14469},
}

@inproceedings{chen2025impress,
      title={IMPRESS: An Importance-Informed Multi-Tier Prefix KV Storage System for Large Language Model Inference},
      author={Weijian Chen and Shuibing He and Haiquan Qu and Rui Zhang and Sheng Yang},
      year={2025},
      booktitle={Proceedings of the 23rd USENIX Conference on File and Storage Technologies (FAST)},
}

@misc{sheng2023flexgen,
      title={FlexGen: High-Throughput Generative Inference of Large Language Models with a Single GPU},
      author={Ying Sheng and Lianmin Zheng and Binhang Yuan and Zhuohan Li and Max Ryabinin and Daniel Y. Fu and Zhiqiang Xie and Beidi Chen and Clark Barrett and Joseph E. Gonzalez and Percy Liang and Christopher Re and Ion Stoica and Ce Zhang},
      year={2023},
      eprint={2303.06865},
      archivePrefix={arXiv},
      primaryClass={cs.LG},
      url={https://arxiv.org/abs/2303.06865},
}

@misc{lee2024infinigen,
      title={InfiniGen: Efficient Generative Inference of Large Language Models with Dynamic KV Cache Management},
      author={Wonbeom Lee and Jungi Lee and Junghwan Seo and Jaewoong Sim},
      year={2024},
      eprint={2406.19707},
      archivePrefix={arXiv},
      primaryClass={cs.LG},
      url={https://arxiv.org/abs/2406.19707},
}

@misc{zheng2026pbkv,
      title={Efficient Serving for Dynamic Agent Workflows with Prediction-based KV-Cache Management},
      author={Haoyu Zheng and Fangcheng Fu and Jia Wu and Binhang Yuan and Yongqiang Zhang and Hao Wang and Yuanyuan Zhu and Xiao Yan and Jiawei Jiang},
      year={2026},
      eprint={2605.06472},
      archivePrefix={arXiv},
      primaryClass={cs.DC},
      url={https://arxiv.org/abs/2605.06472},
}

@inproceedings{lin2024parrot,
      title={Parrot: Efficient Serving of LLM-based Applications with Semantic Variable},
      author={Chaofan Lin and Zhenhua Han and Chengruidong Zhang and Yuqing Yang and Fan Yang and Chen Chen and Lili Qiu},
      year={2024},
      booktitle={Proceedings of the 18th USENIX Symposium on Operating Systems Design and Implementation (OSDI)},
      eprint={2405.19888},
      archivePrefix={arXiv},
      url={https://arxiv.org/abs/2405.19888},
}

@misc{lin2024infinitellm,
      title={Infinite-LLM: Efficient LLM Service for Long Context with DistAttention and Distributed KVCache},
      author={Bin Lin and Chen Zhang and Tao Peng and Hanyu Zhao and Wencong Xiao and Minmin Sun and Anmin Liu and Zhipeng Zhang and Lanbo Li and Xiafei Qiu and Shen Li and Zhigang Ji and Tao Xie and Yong Li and Wei Lin},
      year={2024},
      eprint={2401.02669},
      archivePrefix={arXiv},
      primaryClass={cs.DC},
      url={https://arxiv.org/abs/2401.02669},
}

@misc{nixon2026yearllmserving,
      title={A Year in {LLM} Serving: Workload Evolution, Caching and Load-Balancing},
      author={William Nixon and Jon Durbin and Florian Standhartinger and Haryadi S. Gunawi and Juncheng Yang},
      year={2026},
      eprint={2608.13573},
      archivePrefix={arXiv},
      url={https://arxiv.org/abs/2608.13573},
}

@misc{chang2026agentsysbench,
      title={From {LLM} Inference to Agentic Workloads: Characterization and Implications for Serving Systems},
      author={Chaokun Chang and Yukun Zhou and Kaihua Fu and Dakai An and Tianyu Feng and Hanfeng Lu and Sheng Yao and Pu Guo and Yinghao Yu and Yizhou Shan and Bo Li and Binhang Yuan and Wei Wang},
      year={2026},
      eprint={2608.15127},
      archivePrefix={arXiv},
      url={https://arxiv.org/abs/2608.15127},
}

@misc{chen2025browsecompplus,
      title={BrowseComp-Plus: A More Fair and Transparent Evaluation Benchmark of Deep-Research Agent},
      author={Zijian Chen and Xueguang Ma and Shengyao Zhuang and Ping Nie and Kai Zou and Andrew Liu and Joshua
Green and Kshama Patel and Ruoxi Meng and Mingyi Su and Sahel Sharifymoghaddam and Yanxi Li and Haoran Hong and X
inyu Shi and Xuye Liu and Nandan Thakur and Crystina Zhang and Luyu Gao and Wenhu Chen and Jimmy Lin},
      year={2025},
      eprint={2508.06600},
      archivePrefix={arXiv},
      primaryClass={cs.CL},
      url={https://arxiv.org/abs/2508.06600},
}

@inproceedings{ren2025agserve,
  author = {Yanyu Ren and Li Chen and Dan Li and Xizheng Wang and
            Zhiyuan Wu and Yukai Miao and Yu Bai},
  title = {Transcending Cost-Quality Tradeoff in Agent Serving via
           Session-Awareness},
  booktitle = {Advances in Neural Information Processing Systems 38
               (NeurIPS 2025)},
  year = {2025},
  doi = {10.52202/085713-0097},
}

\clearpage
\appendix
\nobalance

\section{Appendix}

\subsection{Trace Collection}
\label{sec:appendix-traces}

Table~\ref{tab:traces} summarizes the traces used in this paper, including their workload type, source, scale, and timestamp origin. \emph{Sess.} and \emph{Reqs.} report the number of sessions and model invocations, \emph{Turns} is the mean number of invocations per session, and \emph{Input} is the mean prompt length in tokens.

The Exgentic traces combine runs from four to five models; their timestamps come
from agent execution and therefore include real tool latency.
RAGPulse uses 512-token blocks rather than the
16-token blocks used by the other block-level traces; its byte-based volume
measurements remain comparable, but its reuse measurements can hide shared
prefixes shorter than one block. RAGPulse reconstructs its context independently
at each turn and therefore has no cumulative session growth. 

\begin{table*}[t]
\centering
\footnotesize
\setlength{\tabcolsep}{4pt}
\renewcommand{\arraystretch}{1.05}
\begin{tabularx}{\textwidth}{
    @{}
    >{\raggedright\arraybackslash}X
    >{\raggedright\arraybackslash}p{0.155\textwidth}
    >{\raggedright\arraybackslash}p{0.085\textwidth}
    >{\raggedleft\arraybackslash}p{0.055\textwidth}
    >{\raggedleft\arraybackslash}p{0.055\textwidth}
    >{\raggedleft\arraybackslash}p{0.045\textwidth}
    >{\raggedleft\arraybackslash}p{0.045\textwidth}
    >{\raggedright\arraybackslash}p{0.12\textwidth}
    @{}
}
\toprule
\textbf{Trace} & \textbf{Workload type} & \textbf{Source} &
\textbf{Sess.} & \textbf{Reqs.} & \textbf{Turns} & \textbf{Input} &
\textbf{Timestamps} \\
\midrule

\multicolumn{8}{@{}l}{\textbf{Raw text}\enspace\emph{Prompts and responses; 16-token blocks reconstructed with the Qwen3-8B tokenizer; semantic-role labels}} \\
\addlinespace[2pt]
WildChat~\cite{zhao2024wildchat}                         & Open-ended chat    & Production & 838K  & 1.96M & 2.3  & 1.4K & Real, 13 months \\
AppWorld$^{*}$~\cite{trivedi2024appworld}                & Tool-using agent   & Benchmark  & 1,500 & 48K   & 32.3 & 83K  & Agent run \\
BrowseComp+$^{*}$~\cite{chen2025browsecompplus}          & Tool-using agent   & Benchmark  & 1,948 & 38K   & 19.7 & 44K  & Agent run \\
SWE-bench$^{*}$~\cite{jimenez2024swebench}               & Coding agent       & Benchmark  & 1,959 & 92K   & 46.8 & 22K  & Agent run \\
$\tau^2$-bench$^{*}$~\cite{yao2024taubench}              &                    &            &       &       &      &      & \\
\hspace{1em}Airline                                      & Tool-using agent   & Benchmark  & 957   & 11K   & 11.4 & 11K  & Agent run \\
\hspace{1em}Retail                                       & Tool-using agent   & Benchmark  & 1,848 & 22K   & 11.9 & 11K  & Agent run \\
\hspace{1em}Telecom                                      & Tool-using agent   & Benchmark  & 1,844 & 30K   & 16.3 & 15K  & Agent run \\
RAGBench~\cite{friel2024ragbench}                        & RAG                & Benchmark  & 67K   & 67K   & 1.0  & 1.1K & Synthetic \\
Open Deep Research~\cite{langchain_open_deep_research}   & Multi-lineage      & Ours & 50 & 1,973 & 39.5 & 6.1K & Agent run, 5.6\,h \\
\addlinespace

\multicolumn{8}{@{}l}{\textbf{Pre-hashed}\enspace\emph{Released block identifiers, 16-token (Qwen-Bailian) or 512-token (RAGPulse); no text}} \\
\addlinespace[2pt]
Qwen-Bailian~\cite{wang2025kvcachewild}                  &                    &            &      &      &     &      & \\
\hspace{1em}To-C                                         & Open-ended chat    & Production & 23K  & 43K  & 1.9 & 2.3K & Real, 2\,h \\
\hspace{1em}To-B                                         & API serving        & Production & 173K & 173K & 1.0 & 0.9K & Real, 2\,h \\
\hspace{1em}Code                                         & Coding agent       & Production & 26K  & 43K  & 1.6 & 5.7K & Real, 2\,h \\
RAGPulse~\cite{wang2025ragpulse}                         & RAG                & Production & 19K  & 35K  & 1.9 & 3.6K & Real, 42 days \\
\addlinespace

\multicolumn{8}{@{}l}{\textbf{Metadata only}\enspace\emph{Timestamps, token counts, and session structure; no block identities}} \\
\addlinespace[2pt]
TraceLab~\cite{zhu2026tracelabcharacterizingcodingagent} & Coding agent       & Production & 4,281 & 357K & 83.4 & 154K & Real, 8 months \\
BurstGPT v2~\cite{wang2024burstgpt}                      & Open-ended chat    & Production & 55K   & 232K & 4.2  & 1.0K & Real, 110 days \\
\bottomrule
\end{tabularx}
\caption{The 11 traces and their 15 workloads, grouped by evidence class
(\S\ref{sec:workloads-method}). $^{*}$Agent runs released by
Exgentic~\cite{bandel2026exgentic}.}
\label{tab:traces}
\end{table*}

\stitle{Metadata-only traces.} TraceLab and BurstGPT contain no prompt text or block identities. TraceLab provides, for each model call, its session, its position in the session (round index), start and end times, and input and output token counts. BurstGPT provides the same for the release window that includes session identifiers, except that it has no turn index, so we order a session's calls by arrival time; calls in a session almost never overlap and never share a start time. We therefore use these traces only for results that depend on timing and context size, and project KV state from token counts.

Where an analysis needs the state a call reuses from its own session, 
we take it to be the previous call's context, 
$\min(\mathit{input}_{k+1}, \mathit{input}_k + \mathit{output}_k)$ tokens, 
since chat and coding agents resend the conversation with each call unless 
they compact it. TraceLab also reports cached-prefix tokens, but these are 
the providers' prompt-cache hits and depend on the providers' caching 
policies, so they cannot serve as a policy-independent measure; 
for consecutive calls less than a minute apart (88\% of calls), 
they agree with our estimate within 2\%.

\stitle{Open Deep Research (ODR) collection.} 
We collected the ODR trace ourselves. Open Deep Research~\cite{langchain_open_deep_research} is a LangGraph application in which a supervisor writes a research brief, delegates subtopics to parallel researcher agents, and then produces a final report from their compressed findings. We run it on 50 English tasks (IDs 51--100) from DeepResearch Bench~\cite{du2025deepresearchbench}, one task at a time, with user clarification disabled. All model roles use Qwen3-8B~\cite{qwen3technicalreport} in bf16, served by vLLM~\cite{kwon2023vllm} on one NVIDIA A100 GPU with a 32,768-token context window, while web search uses Tavily. To fit within this context window, we reduce several upstream limits, including the number of concurrent researchers, supervisor rounds, tool-calling rounds, and output lengths. Our goal is not to reproduce the benchmark configuration or evaluate task quality, but to study the resulting KV-access patterns. These changes therefore affect the scale of individual executions rather than the role of the trace in our analysis of multi-lineage context and reuse behavior.

We record every model call at the client's HTTP transport, capturing the exact request messages, response, and wall-clock start and end times of each chat-completion call, together with its task identifier. Because tasks run sequentially, the timestamps capture concurrency among researchers within a task but not across concurrent users. The trace contains 1,973 model invocations over 5.6 hours. Unlike a chat session, a single task contains many independent message histories: the supervisor, each researcher, and the report writer each maintain their own. We recover these histories, which we call \emph{lineages}, from the messages themselves. A call continues a lineage if the messages and response of an earlier call appear unchanged at the start of its input; when multiple candidates match, we link it to the longest one. A call with no match starts a new lineage, such as a newly launched researcher. Each task contains about 33 lineages (\S\ref{sec:reuse}). We then tokenize and block the calls using the same procedure as for the other raw-text traces (\S\ref{sec:workloads-method}).

\begin{table*}[t]
\centering
\footnotesize
\setlength{\tabcolsep}{3.5pt}
\renewcommand{\arraystretch}{1.05}
\begin{tabular}{@{}l rr rrr r r rr rrr@{}}
\toprule
& \multicolumn{2}{c}{\textbf{Turns}}
& \multicolumn{3}{c}{\textbf{Growth}}
& \multicolumn{2}{c}{\textbf{Single-session}}
& \multicolumn{2}{c}{\textbf{Multi-session}}
& \multicolumn{3}{c}{\textbf{Multi-session tail}} \\
& & & & & & \multicolumn{2}{c}{\textbf{state (GiB)}} & \multicolumn{2}{c}{\textbf{state (GiB)}} & \multicolumn{3}{c}{\textbf{(\%)}} \\
\cmidrule(lr){2-3} \cmidrule(lr){4-6} \cmidrule(lr){7-8} \cmidrule(lr){9-10} \cmidrule(l){11-13}
\textbf{Workload}
& p50 & p99
& \textbf{Start} & \textbf{MiB/} & $R^2$
& \textbf{Peak} & \textbf{Reused}
& \textbf{All} & \textbf{50\% of}
& \textbf{Mem.} & \textbf{Mem.} & \textbf{Hits} \\
& & & \textbf{(GiB)} & \textbf{turn} &
& \textbf{p99} & \textbf{p50}
& & \textbf{hits}
& $k{=}2$ & $k{\le}5$ & \textbf{lost}  \\
\midrule
\multicolumn{13}{@{}l}{\emph{Single-request}} \\
Qwen To-B & 1 & 1 & -- & -- & -- & 0.95 & -- & 479 & 2.43 & 45 & 71 & 4 \\
RAGBench & 1 & 1 & -- & -- & -- & 1.91 & -- & 2,407 & 443 & 59 & 91 & 63 \\
\addlinespace
\multicolumn{13}{@{}l}{\emph{Short multi-turn}} \\
Qwen Code & 1 & 9 & 0.58 & 79 & 0.99 & 2.49 & 0.40 & 1,114 & 2.86 & 51 & 86 & 13 \\
Qwen To-C & 1 & 8 & 0.11 & 72 & 1.00 & 2.06 & 0.16 & 113 & 0.09 & 59 & 87 & 10 \\
RAGPulse & 1 & 9 & -- & -- & -- & 1.46 & 0.14 & 30.0 & 0.14 & 77 & 80 & 1 \\
WildChat & 1 & 14 & 0.01 & 32 & 1.00 & 1.43 & 0.07 & 4,496 & 254 & 55 & 85 & 35 \\
\addlinespace
\multicolumn{13}{@{}l}{\emph{Agents}} \\
$\tau^2$-Airline & 10 & 50 & 0.67 & 78 & 0.98 & 4.87 & 1.35 & 35.6 & 8.68 & 6 & 21 & 2 \\
$\tau^2$-Retail & 11 & 51 & 2.60 & 28 & 0.95 & 4.35 & 1.48 & 36.5 & 8.27 & 10 & 23 & 1 \\
$\tau^2$-Telecom & 13 & 105 & 3.14 & 11 & 0.99 & 4.41 & 1.55 & 140 & 12.4 & 31 & 60 & 6 \\
BrowseComp+ & 13 & 96 & 1.71 & 222 & 1.00 & 24.7 & 4.62 & 18.4 & 5.18 & 1 & 9 & 0 \\
AppWorld & 19 & 150 & 12.1 & 60 & 0.97 & 31.2 & 12.7 & 838 & 49.9 & 66 & 82 & 10 \\
SWE-bench & 38 & 159 & 0.67 & 71 & 0.99 & 13.1 & 3.02 & 30.3 & 6.10 & 9 & 15 & 1 \\
TraceLab & 17 & 1,162 & 4.80 & 219 & 0.96 & 58.6 & -- & -- & -- & -- & -- & -- \\
\bottomrule
\end{tabular}
\caption{Per-workload volume and footprint under the Qwen3-8B bf16 KV geometry (144\,KiB per token).}
\label{tab:volume-summary}
\end{table*}

\begin{figure}[t]
    \centering
    \includegraphics[width=\linewidth]{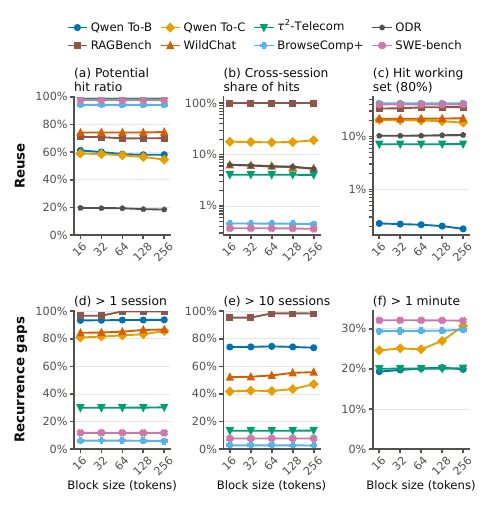}
    \caption{Reuse (top) and cross-session recurrence (bottom) for block sizes of 16--256 tokens.}
    \label{fig:block-sensitivity}
\end{figure}

\begin{figure}[t]
    \centering
    \includegraphics[width=\linewidth]{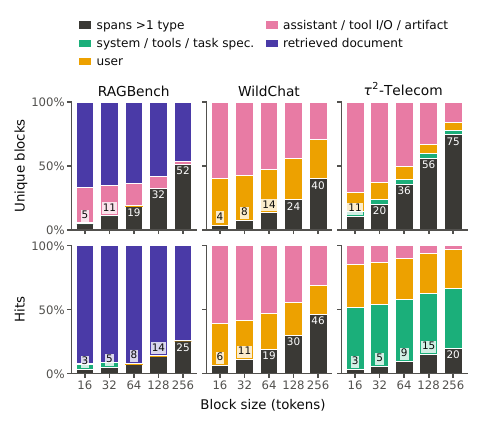}
    \caption{Context-type composition of unique blocks (top) and hits (bottom) for block sizes of 16--256 tokens. Black: blocks whose tokens span more than one context type; the number gives their share in percent.}
    \label{fig:block-roles}
\end{figure}

\subsection{Block Sensitivity}
\label{sec:block-sensitivity}

All results in \S\ref{sec:analysis} use 16-token blocks. To test whether the findings depend on this choice, we repeat the reuse and temporal measurements with 32-, 64-, 128-, and 256-token blocks on eight workloads, including Qwen To-C and BrowseComp+, which anchor the temporal results. Because block identifiers are prefix-chained, a $b$-token block is identified by the last of its $b/16$ 16-token sub-blocks, so we derive the larger block sizes from the 16-token traces rather than re-tokenizing.

\stitle{Reuse.} Block size does not change the reuse picture (Figure~\ref{fig:block-sensitivity}a--c). The ordering of workloads is preserved in every panel, except that Open Deep Research and WildChat, whose cross-session shares are both 6.3\% at 16 tokens, trade places. The agent workloads and WildChat keep the same hit ratio at every block size. Hit ratios decline slightly only in Qwen To-C (59\% to 55\% at 256 tokens), Qwen To-B (61\% to 58\%), RAGBench (71\% to 70\%), and Open Deep Research (19.7\% to 18.5\%). Cross-session shares move by at most 1.1 percentage points and hit working sets by at most 2.2 points. Larger blocks mainly reduce reuse near prefix boundaries: when two sequences diverge within a block, the shared tokens in that partially matching block can no longer be reused. This matters for short shared prefixes, but has little effect when reuse spans thousands of tokens, as in the agent workloads.

\stitle{Recurrence.} Recurrence gaps are equally stable (Figure~\ref{fig:block-sensitivity}d--f). In the agent workloads, every share stays within 0.6 percentage points. The short-context workloads move by at most 6 points, mostly upward. RAGBench shows why: every request begins with the same 32--47-token instruction prefix, which fills two 16-token blocks and one 32-token block but no larger block. From 64 tokens onward, this block, which every request shares, disappears, and the share of gaps longer than one session rises from 97\% to nearly 100\%. Qwen To-C and WildChat drift more gradually, consistent with the same loss of short shared prefixes: Qwen To-C's share of gaps above ten sessions rises from 42\% to 47\% and above one minute from 25\% to 31\%.

\stitle{Context-type composition.} The semantic role of a block is where block size matters most (Figure~\ref{fig:block-roles}). Roles change at turn and message boundaries, so larger blocks increasingly span more than one context type, and their share grows with every doubling of block size. At 16 tokens, such blocks are 4--11\% of unique blocks and supply at most 6\% of hits, so the role attribution in \S\ref{sec:reuse} is reliable. At 256 tokens, they are 40\% of unique blocks in WildChat, 52\% in RAGBench, and 75\% in $\tau^2$-Telecom. Their share of hits stays lower in RAGBench and $\tau^2$-Telecom, 25\% and 20\% at 256 tokens, because much of the reuse comes from long single-type runs such as retrieved documents and tool definitions; in WildChat, it reaches 46\%. A role label per block is therefore a relatively reliable admission hint for small blocks but becomes ambiguous for large ones.

\stitle{Volume.} We do not repeat the volume analysis: KV bytes are proportional to tokens, and the context-growth fits of Figure~\ref{fig:volume} are computed from token counts, so block size affects shared volume only through the short-prefix effect described above.

\storageimplication{Block size does not change the conclusions of \S\ref{sec:analysis}. It matters in two places. Shared prefixes shorter than a block are invisible to storage, so blocks used for cross-session sharing should be no larger than the shortest stable prefix a workload repeats. Role labels lose precision as blocks grow, so role-based hints need small blocks or token-range metadata. Deep private session state can use large blocks, which cut the number of block entries by $16\times$ from 16 to 256 tokens without losing reuse.}

\subsection{Volume Summary}
\label{sec:appendix-volume}

Table~\ref{tab:volume-summary} reports the volume metrics of \S\ref{sec:volume}, grouped into single-request workloads, short multi-turn workloads, and agents. \emph{Turns} reports the median and 99th-percentile number of invocations per session. \emph{Growth} fits a least-squares line to the median context size at each turn among sessions that reach the same depth: \emph{Start} is the intercept, \emph{MiB/turn} is the slope, and $R^2$ reports goodness of fit.

For single-session state, \emph{Peak} is the 99th percentile of each session's maximum context size, and \emph{Reused} is the median amount of that session's own KV state needed to supply all of its intra-session hits. Because this state is private to a session, each concurrently live session incurs this footprint separately.

For multi-session state, \emph{All} is the total size of blocks reused by more than one session. \emph{50\% of hits} reports the amount of this state needed to supply half of all cross-session hits when blocks are ranked by reuse frequency. The \emph{tail} columns characterize low-frequency shared state: they report the fraction of multi-session memory occupied by blocks used by exactly two sessions ($k{=}2$) or by at most five sessions ($k{\le}5$), together with the fraction of cross-session hits lost if blocks used by at most five sessions are discarded.

\subsection{Recurrence Gaps in Sessions}
\label{sec:appendix-recurrence}

\begin{figure}[t]
    \centering
    \includegraphics[width=1\linewidth]{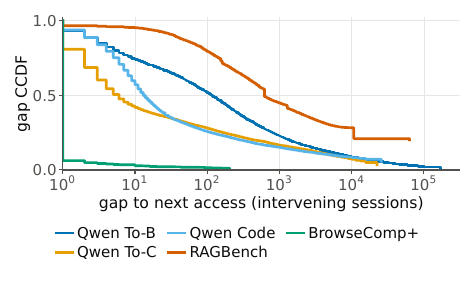}
    \caption{Recurrence gaps of multi-session state, counted in session start
    order: a gap of one means the next session to start reuses the block.}
    \label{fig:appendix-recurrence}
\end{figure}

Figure~\ref{fig:appendix-recurrence} measures the gaps of
Figure~\ref{fig:temporal-ttl}c in intervening sessions rather than seconds.
RAGBench has no real timestamps, so it is absent from
Figure~\ref{fig:temporal-ttl}c but included here.
BrowseComp+'s common prefix recurs almost every session: only 6\% of gaps
exceed one session.
In Qwen To-B, half of the gaps exceed 100 sessions and 9\% exceed 10,000; in
RAGBench, 79\% exceed 100 sessions and 45\% exceed 1,000.

\subsection{Fixed-Concurrency Replay}
\label{sec:appendix-temporal}
Figure~\ref{fig:temporal-concurrency} keeps exactly $N$ sessions live at all times. 
Each of $N$ slots replays randomly chosen recorded sessions from the trace back to back, 
starting a new one as soon as the previous one ends. Each session keeps its recorded call 
times and context sizes. A session is live from the start of its first call to the end 
of its last. We convert tokens to bytes at 144\,KiB per token, the bf16 KV size of 
Qwen3-8B. The Exgentic panel draws sessions from all six Exgentic workloads
(AppWorld, BrowseComp+, SWE-bench, and the three $\tau^2$-bench domains).
In Exgentic, blocks shared with other sessions are counted once while any 
session that uses them is live. BurstGPT and TraceLab expose no block identities, 
so all of their state is counted as private. The dashed line instead counts each 
live session's full context. Each bar reports the mean resident state once the replay 
reaches a steady state.

\subsection{Session Completion and Shared State}
\label{sec:appendix-completion}

\begin{figure}[t]
    \centering
    \includegraphics[width=1\linewidth]{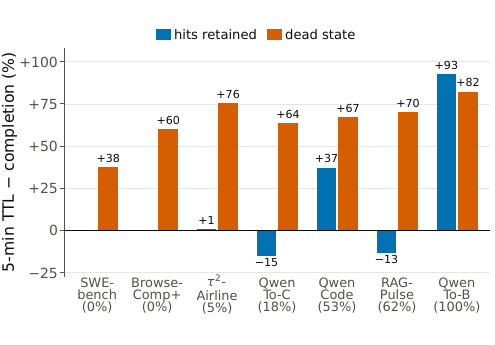}
    \caption{A five-minute idle
    TTL minus reclamation at session completion, for all hits retained
    (within- and cross-session) and dead state. Brackets: share of re-reads
    that are cross-session.}
    \label{fig:appendix-completion}
\end{figure}

Figure~\ref{fig:info}b considers only within-session hits, for which
reclamation at completion is lossless. Figure~\ref{fig:appendix-completion}
extends the replay to include cross-session reuse. At a session's final request,
completion frees every block touched by that session unless another live
session still holds it; the TTL instead frees blocks after five minutes of
inactivity. Dead state is the share of resident block-seconds occupied by
blocks whose holding sessions have all completed and that are never accessed
again.

When re-reads are almost entirely within-session (SWE-bench, BrowseComp+,
$\tau^2$-Airline), completion preserves within one percentage point of the TTL's
hits while holding 38--76\% less dead state. In Qwen To-C and
RAGPulse, it retains 15\% and 13\% more hits because the TTL prematurely expires paused
sessions. Completion instead loses hits when shared blocks recur only after
every session holding them has completed. RAGPulse's shared blocks recur quickly: only
6\% of their recurrence gaps exceed ten sessions, so a session holding them is
usually still live. In Qwen Code, 57\% exceed ten sessions
(Figure~\ref{fig:appendix-recurrence}), and the TTL retains
37\% more of all hits. Qwen To-B sessions
are single requests, 
so every block is freed when its only request ends; all of
its hits are cross-session, 
and the TTL retains 93\% more of them.

This loss does not come from the lifecycle signal itself but from applying it
to multi-session state, which has no completion boundary
(\S\ref{sec:temporal}). With a session ID, storage knows which blocks have been
touched by more than one session (\S\ref{sec:analysis-interface}) and can
exclude them from reclamation, applying lifecycle only to single-session state.
The shared blocks it spares should then be retained according to their
recurrence and capacity cost (\S\ref{sec:reuse}) rather than their sessions'
lifecycle.

\end{document}